\documentclass[conference]{IEEEtran}
\usepackage{url}
\usepackage{amsmath}
\usepackage[pdftex]{graphicx}
\usepackage{epstopdf}

\usepackage{caption}
\usepackage{subcaption}

\usepackage[linesnumbered,ruled]{algorithm2e}

\IEEEoverridecommandlockouts

\begin{document}
	
	\title{Empirical Study of Drone Sound Detection in Real-Life Environment with Deep Neural Networks}
	
	
	
	\author{\IEEEauthorblockN{Sungho Jeon\textsuperscript{1}, Jong-Woo Shin, Young-Jun Lee, Woong-Hee Kim, YoungHyoun Kwon, and Hae-Yong Yang}
	\IEEEauthorblockA{The Affiliated Institute of ETRI \\
	Daejeon, South Korea\\
	Email: \{sdeva, jwshin, lhyjlee, whkim, wishwill, formant\}@nsr.re.kr}
	}
	
	\maketitle
	
	
	

	\begin{abstract}
		
		This work aims to investigate the use of deep neural network to detect commercial hobby drones in real-life environments by analyzing their sound data. The purpose of work is to contribute to a system for detecting drones used for malicious purposes, such as for terrorism. Specifically, we present a method capable of detecting the presence of commercial hobby drones as a binary classification problem based on sound event detection. We recorded the sound produced by a few popular commercial hobby drones, and then augmented this data with diverse environmental sound data to remedy the scarcity of drone sound data in diverse environments. We investigated the effectiveness of state-of-the-art event sound classification methods, i.e., a Gaussian Mixture Model (GMM), Convolutional Neural Network (CNN), and Recurrent Neural Network (RNN), for drone sound detection. Our empirical results, which were obtained with a testing dataset collected on an urban street, confirmed the effectiveness of these models for operating in a real environment. In summary, our RNN models showed the best detection performance with an F-Score of 0.8009 with 240 ms of input audio with a short processing time, indicating their applicability to real-time detection systems.
		
	\end{abstract}
	
	\IEEEpeerreviewmaketitle

	\footnotetext[1]{Contact email: sdeva14@gmail.com}

	\section{Introduction}
	\textbf{Motivation.} Popularization of commercial hobby drones brings unexpected threats to the environment in which we live, such as terror to people or important facilities. A common four-propeller drone is suitable to enjoy as a hobby and for broadcasting, however, at the same time, it surprisingly makes existing defense systems appear to be outdated legacy systems. Some accidents already proved that these drones can easily penetrate the highest level of security systems, such as landing in front of the prime minister of Germany, on the rooftop of the official residence of the prime minister of Japan, and at the White House in the United States. Thus, the ability to detect the appearance of a drone is a matter of the highest priority to prevent any threats. 
	
	\textbf{Existing work.} Even though few studies have been concerned with the problem of drone sound detection, previous work was conducted in isolated or calm places rather than in a real-life environment without the polyphonic sound environment typical of outside areas, such as on the rooftop of a building in a calm place or isolated environment \cite{mezei2015drone, busset2015detection, mezei2016drone}. However, considering our target problem, which is to detect drones used for malicious purposes, the system inevitably needs to be utilized in a real-life environment, and this requires us to consider polyphonic sound data. 
  Other work differs by using an impressive approach based on radar information or the RF frequency \cite{mendis2016deep, nguyen2016investigating}, but we need to consider a combined detection system with a multiple approach to complement the drawback of each method.
	
	Event Sound Classification (ESC) in a real environment has been highlighted for diverse purposes. Many researchers have focused on finding useful features and classifiers based on the machine-learning approach. The most popular combination of feature and classification is Mel-frequency Cepstrum Coefficients (MFCC) \cite{ barchiesi2015acoustic} with the Gaussian Mixture Model (GMM) \cite{pohjalainen2011detection, mesaros2016tut}. 
  More recently, the impressive success achieved with Deep Neural Networks (DNNs) has motivated researchers to introduce these networks to environmental sound recognition. Two popular DNN models, the Convolutional Neural Network (CNN) \cite{zhang2015robust, cakir2016filterbank} and Recurrent Neural Network (RNN) \cite{parascandolo2016recurrent}, have also been highlighted for audio-related tasks. Even though these previous studies cover the ESC problem, considering the importance and urgency of our problem in terms of terrorism, it is worth exploring how ESC work can be applied and to assess its effectiveness for drone sound detection. Here it should be noted that rather than intended to propose novel features or models for drone sound detection, our work aims to investigate the practical effectiveness of popular classification models for our problem in real environments used in previous ESC studies.
	
	
	
	\textbf{Contribution.} Our contributions are summarized as follows:
	\begin{itemize}
		\item  To the best of our knowledge, we are the first to investigate drone sound detection in highly noisy real environments with the aim of constructing a detection method for practical usage with real-time systems based on three popular ESC models: GMM, CNN, and RNN.
		\item We show that the shortage of training data for a drone sound classification model can be remedied with our audio augmentation that synthesizes raw drone sound with diverse background sounds.
		\item We investigate the effectiveness of these models for a testing dataset collected from real-life environments in terms of the F-Score and by taking consideration of the processing time for application to real-time systems.
		
	\end{itemize}
	
	
	\section{Method}
	
	\subsection{Data Augmentation} 
  Especially in real environments, unseen event sound  has a detrimental effect in terms of deterioration of the detection rate. The most challenging difficulty for this work is the absence of public drone sound data for training. Even though supplying for commercial hobby drone is available, collecting drone sound in diverse environments is only possible to a limited extent, because flying a drone in most public or residential areas is restricted. We therefore remedied the shortage of training data, by augmenting the drone sound with diverse real-life environmental sounds from a public dataset \cite{Grootel2009, rakotomamonjy2015histogram} and our collection. The drone sounds were collected in a quiet place outside. The purpose of this augmentation is to produce drone sound combined with realistic noise data, while preserving the characteristics of drone sound. Data augmentation involved amplifying the power of the drone sound such that it exceeded that of the background sound data by 5\% in terms of max peak to emphasize the characteristics of the drone sound. Our augmented audio clip consists of concatenated raw background sound and overlapped background sound with repeated drone sounds equal to the length of the background.
	
	\subsection{Feature: MFCC and Mel-spectrogram} In many previous ESC studies, MFCC is known to possess outstanding features for classifiers. MFCC also has useful features to capture periodicity from the fundamental frequencies caused by the rotor blades of a drone. Our recorded drone sound indicated a noticeable harmonic shape below a frequency of 1500 Hz. In addition, we also observed a noticeable influence area on the spectrogram between 5000 Hz and 7000 Hz (Figure \ref{fig:label}) as was previously pointed out \cite{mezei2015drone, busset2015detection}. However, these characteristics were not exhibited by all the drone models used in our experiments. Furthermore, low-frequency data is carried farther than high-frequency data in terms of their energy. Therefore, we only focus on low-frequency data below 1500 Hz. 
	
	The other important consideration for feature engineering is the length of instantaneous input data to the model. The minimum length of audio data converted to an MFCC vector and that shows the best performance with our GMM configuration is 40 ms with 50\% of overlapping. The other models, CNN and RNN, deliver the best performance when they process data of at least 240 ms in length, converted to mel-spectrogram with mel-bin as 40.
	
	
	\subsection{Classifier1: Gaussian Mixture Model} 
	The GMM detector we construct consists two GMMs trained by positive and negative respectively. For a given length of audio data, it is clipped as fixed windows, which is described as the sample $X=x_1,x_2,...,x_l$, where $l$ is the frame length. Then we compare the log-likelihood ($L$) of both models with a decision threshold to decide drone appearance, $Label_{predicted}=L_1 - L_2 > \theta_{decision}$. In our experiment, GMM, with the number of Gaussian as 13, the number of MFCC as 20, and the number of mel-bin as 40, shows the best detection performance. Higher values for these parameters, as proposed in previous work, lead to the overfitting problem that shows higher detection performance in training, but produces dissatisfactory results on the testing dataset. The type of covariance shape affected by the detection performance is nearly 0.1 in our training, although we apply a diagonal shape instead of a full shape to alleviate the over-fitting effect. 
	
	\subsection{Classifier2: Convolutional Neural Network}
\begin{figure*}[ht!]
		\centering
		\begin{subfigure}[b]{0.45\textwidth}
			\includegraphics[scale=0.25]{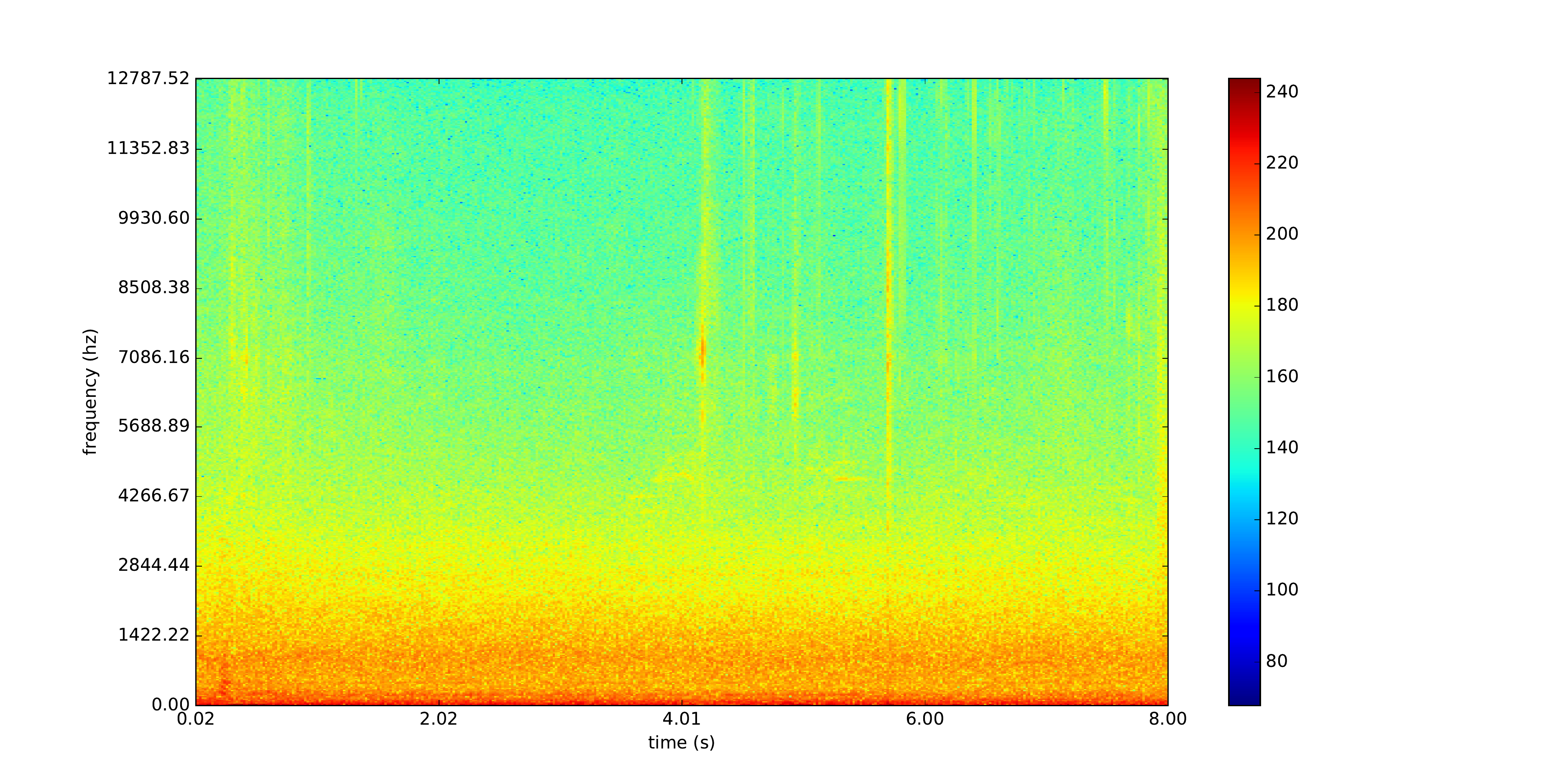}
			\caption{Spectrogram of negative data (Freq: 0$\sim$12k)}
			\label{fig:spectro_neg}
		\end{subfigure}
		\begin{subfigure}[b]{0.45\textwidth}
			\includegraphics[scale=0.25]{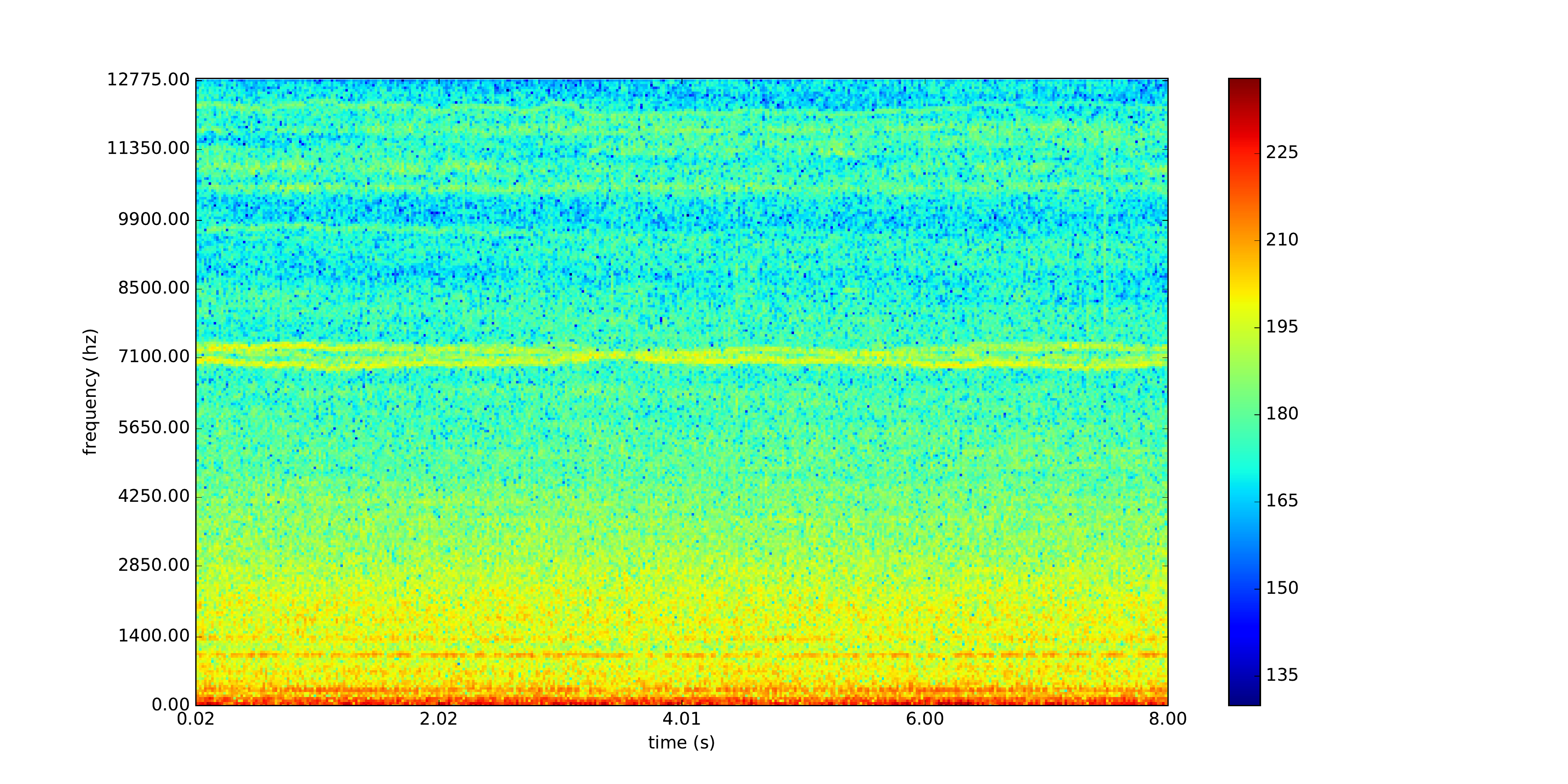}
			\caption{Spectrogram of positive data (Freq: 0$\sim$12k)}
			\label{fig:spectro_pos}
		\end{subfigure}
		\begin{subfigure}[b]{0.45\textwidth}
			\includegraphics[scale=0.135]{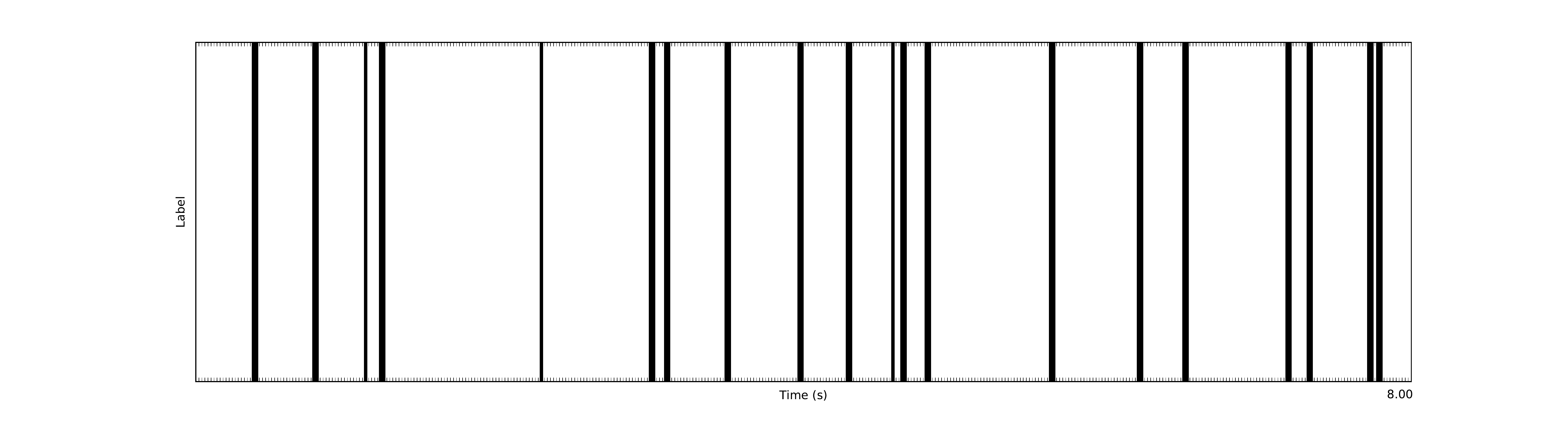}
			\caption{Predicted detection label w.r.t time from GMM}
		\end{subfigure}
		\begin{subfigure}[b]{0.425\textwidth}
			\includegraphics[scale=0.135]{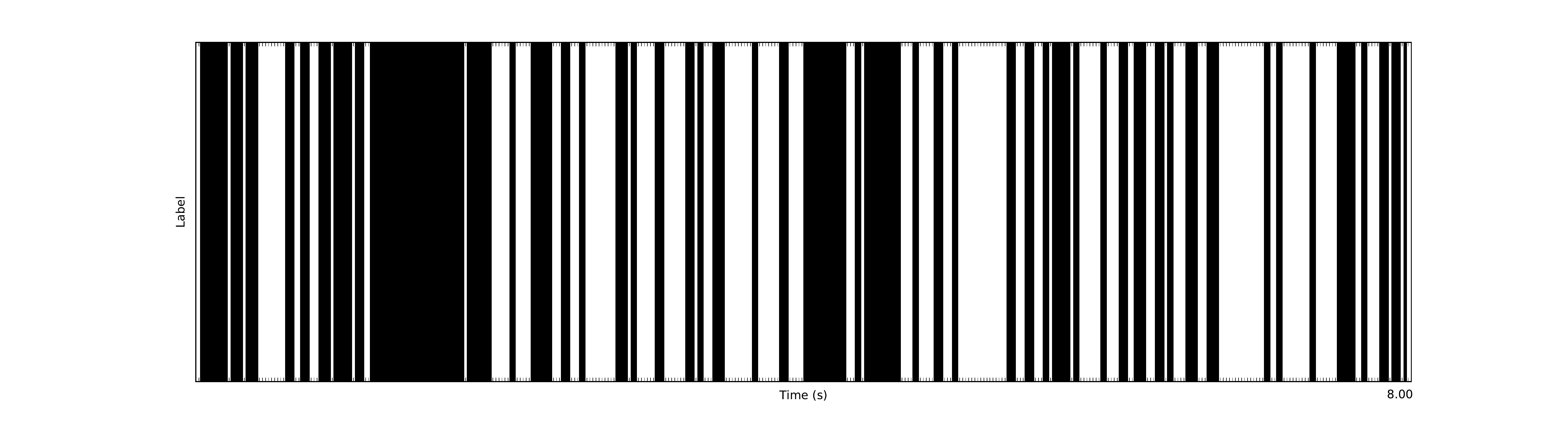}
			\caption{Predicted detection label w.r.t time from GMM}
		\end{subfigure}
		\begin{subfigure}[b]{0.45\textwidth}
			\includegraphics[scale=0.135]{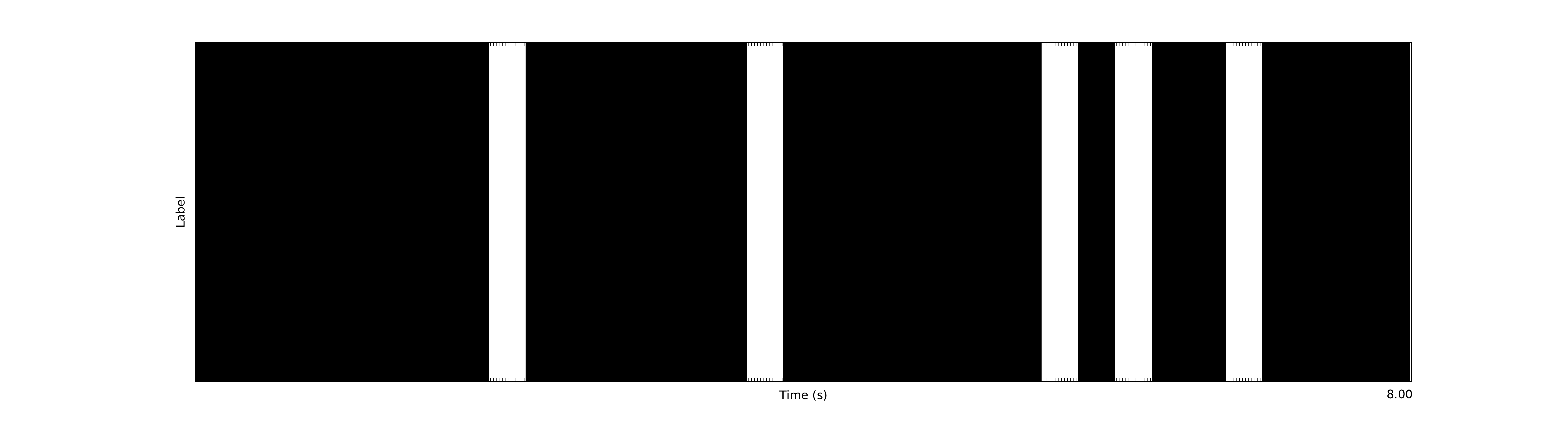}
			\caption{Predicted detection label w.r.t time from CNN}
		\end{subfigure}
		\begin{subfigure}[b]{0.425\textwidth}
			\includegraphics[scale=0.135]{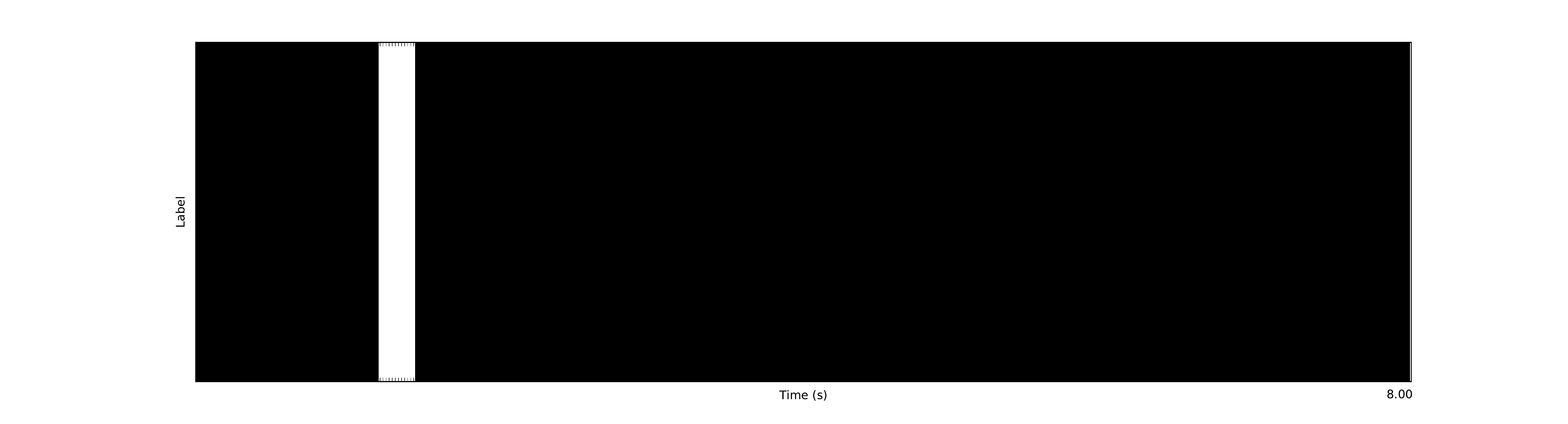}
			\caption{Predicted detection label w.r.t time from CNN}
		\end{subfigure}
		\begin{subfigure}[b]{0.45\textwidth}
			\includegraphics[scale=0.135]{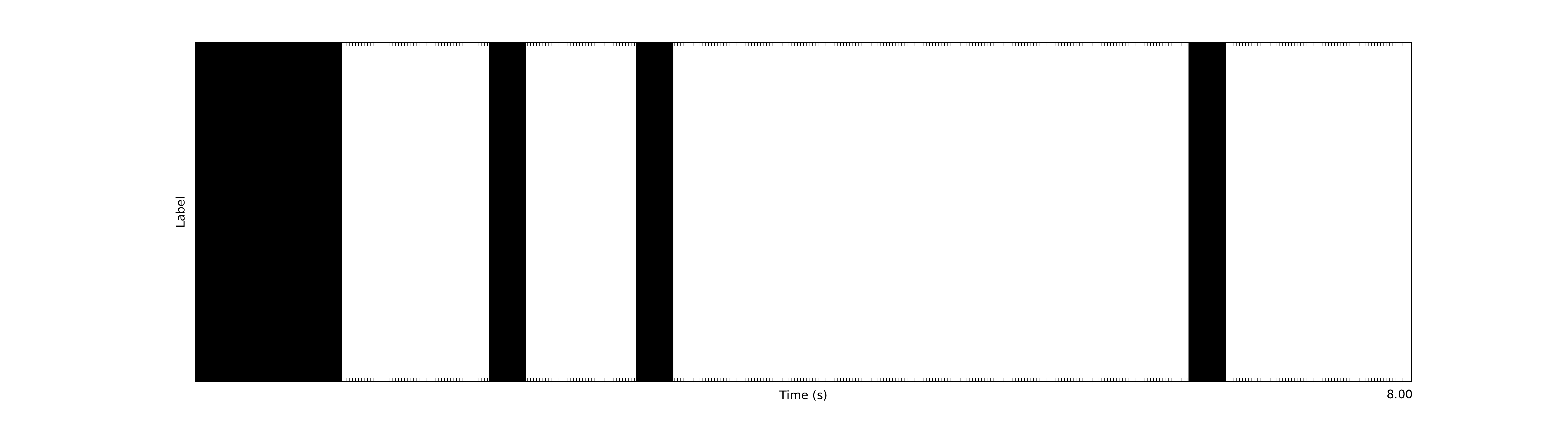}
			\caption{Predicted detection label w.r.t time from RNN}
		\end{subfigure}
		\begin{subfigure}[b]{0.425\textwidth}
			\includegraphics[scale=0.135]{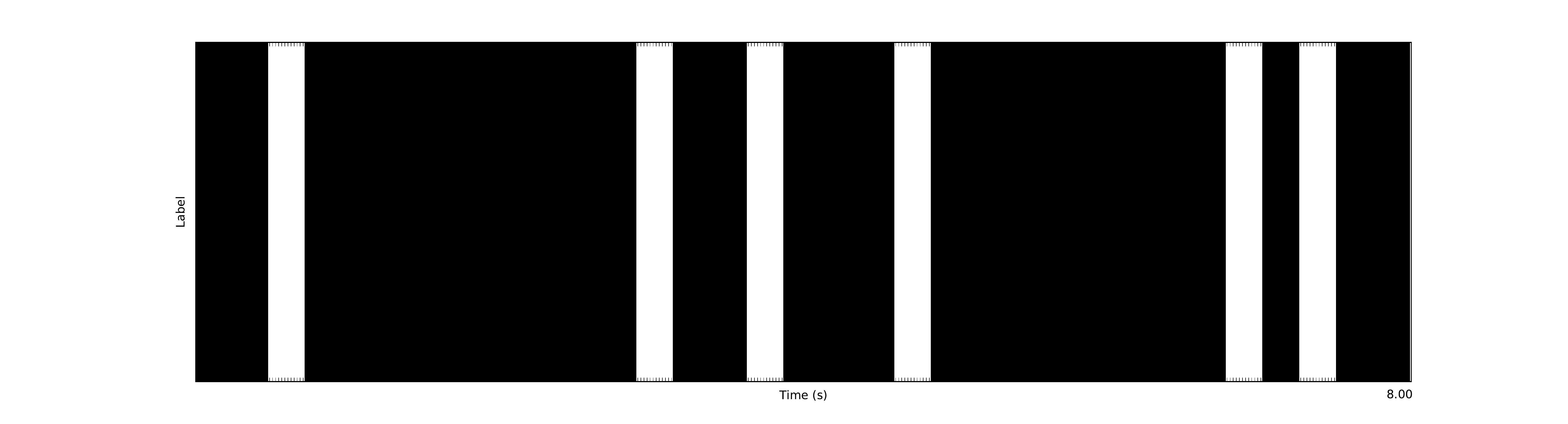}
			\caption{Predicted detection label w.r.t time from RNN}
		\end{subfigure}
		\caption{Example: Spectrogram of negative and positive data and corresponding detection label from urban street (black area in Figure (c)$\sim$(h) indicates predicted period during which drone exists)}
		\label{fig:label}
	\end{figure*}

	CNN for audio-related tasks showed outstanding results with spectral features instead of focusing on feature engineering \cite{zhang2015robust, cakir2016filterbank}. The main idea of a CNN is the use of a convolutional layer that performs localized filtering for local connectivity. This local connectivity is known to be effective to capture invariance useful patterns and highly correlated values with time-frequency representation of sound signal data. Our observation that drone sound has noticeable invariance characteristics below 1500 Hz with harmonics (Figure \ref{fig:label}). 
	
  Our proposed simple architecture consists of nine stages contrary to previous approaches proposed in audio-related tasks (Table \ref{table:cnn_arch}), because rather than improving the performance, a more complex model easily leads to the overfitting problem. During training, we periodically checked the accuracy and loss with the testing dataset, then stopped training if the accuracy did not improve for three epochs of training. Eventually, we selected the model that showed the best accuracy. We shuffled the training dataset every epoch with a learning rate of 0.001 and a batch size of 128. 
	
	\begin{table}
		\caption{Our CNN architecture}
		\label{table:cnn_arch}
		\centering
		\begin{tabular}{|l | l|}
			\hline
			Input size & Description \\ \hline
			(3, 3, 1, 32) & (3, 3) reception field with kernel size 1 to 32 \\
			(3, 3, 32, 32) & (3, 3) reception field with identical kernel size \\
      - & max pooling with (1, 2, 2, 1) \\ 
			Drop-out & Drop-out with 0.5 probability \\
			(3, 3, 32, 256) & (3, 3) reception field with kernel size 32 to 256 \\
			(3, 3, 256, 256) & (3, 3) reception field with identical kernel size \\
      - & max pooling with (1, 2, 2, 1) \\
			Drop-out & Drop-out with 0.5 probability \\
			(3*10*256, 1024) & Full-connected layer \\
			Drop-out & Drop-out with 0.5 probability \\
			(1024, 2) & binary output class label \\
			\hline
		\end{tabular}
	\end{table}
	
	\subsection{Classifier3: Recurrent Neural Network} 
	The other popular DNN model, RNN, is designed to make use of past information to feedforward the network. They perform the same task repeatedly with memory, which represents the context of the information accumulated up to that moment. This memory component has the role of preventing the vanishing gradient problem that decays the influence of past data. Based on this idea, the long short-term memory (LSTM) design is commonly used for standard RNN through replacing simple neurons to LSTM memory blocks, which consist of several gates, such as a $tanh$ input gate, a forget gate to decide whether to remain, and an output gate to control which value is used to compute the output activation. Finally, the output of the LSTM memory block is computed as a multiplication of these gates. 
	
	In this work, bi-directional LSTM-RNN with three layers and 300 LSTM blocks shows the best detection performance. More complicated network configuration shows worse performance or easily results in overfitting. 
	Likewise, when training the CNN model, an early stopping strategy is used in RNN model training by periodically checking the accuracy and loss from the testing dataset. We stop training if it is not improved over 3 epochs of training, after which we retain the model that shows the best accuracy. We shuffle the training dataset every epoch and use a learning rate of 0.0005 and batch size of 64.
	
	\section{Experiment}
	
	We evaluated our methodology by answering the following questions: (Q1) comparing the detection performance of three models, GMM, CNN, and RNN. (Q2) determining the detection performance for unseen types of data such as detecting different drone models or in different environments. (Q3) considering the required computing cost for application to real-time detection systems. All reported performance values are averaged across 10 evaluation results. We implemented the model in Python 2.7 with Scikit-learn 0.18, Librosa 0.4.3, and Tensorflow 0.12 on the following system: 4-core 2.6-GHz CPU, SSD, and GTX 1070.
	
	\subsection{Data description}
	Our training dataset is augmented by raw background sound and drone sound. The background sound consists of data from our own recording and from a public dataset \cite{Grootel2009, rakotomamonjy2015histogram} and the drone sound was collected manually with two popular commercial drones -- Phantom3 and Phantom4 from DJI. Our background sound data contains sounds from ordinary real-life situations with common noise such as chatting, car passing, and airplane noise with a total time of 677 seconds. Our drone sound data was recorded in a quiet outdoor place at a distance of 30m, 70m, and 150m for two types of behavior, hovering and approaching, with a total time of 64 seconds. Exact labeling of the drone sound was achieved by starting to record after the drone is activated, and stopping before deactivation. As a result of augmentation, the total audio time used for training is 9556 seconds.
	
	
	\begin{table}[ht!]
		\caption{Data Description}
		\label{table:eTime}
		\centering
		\begin{tabular}{| l | c | l |}
			\hline
			Data Type & Total time (s) & Description \\ 
			\hline
			Raw: background & 677 & Background audio used for augmentation \\
			Raw: drone & 64 & Drone audio used for augmentation \\
			Training: augmented &  9556.68 & augmented data for training \\
			Testing: detection & 151.06 & measured in urban street for testing \\
			Testing: unseen & 1557 & measured in outside with unseen type data \\
			\hline
		\end{tabular}
	\end{table}
	
	Note that we separate the training and testing datasets to enable us to strictly measure the performance, instead of the k-fold cross validation technique, which is commonly used to remedy a data shortage. Although an augmented dataset is useful for training, it has limited scope for completely mimicking a real dataset. We observe that the real dataset is not completely reproduced by augmentation, due to the complexity of audio characteristics and influence from the environment. Our testing dataset was collected on a real urban street, half of the data relating to a normal situation and the other half in proximity of a building construction site for 151 seconds with an equal amount of positive and negative data. Additionally, we built another testing dataset to measure detection performance for unseen types of data in training, such as unseen types of drones and background. This dataset includes other drone types, DJI Inspire and 3DR Solo, and other types of background such as near a highway or a very noisy road.
	
	\subsection{Testing: detection performance}
\begin{figure}[ht!]
		\centering
		\includegraphics[scale=0.35]{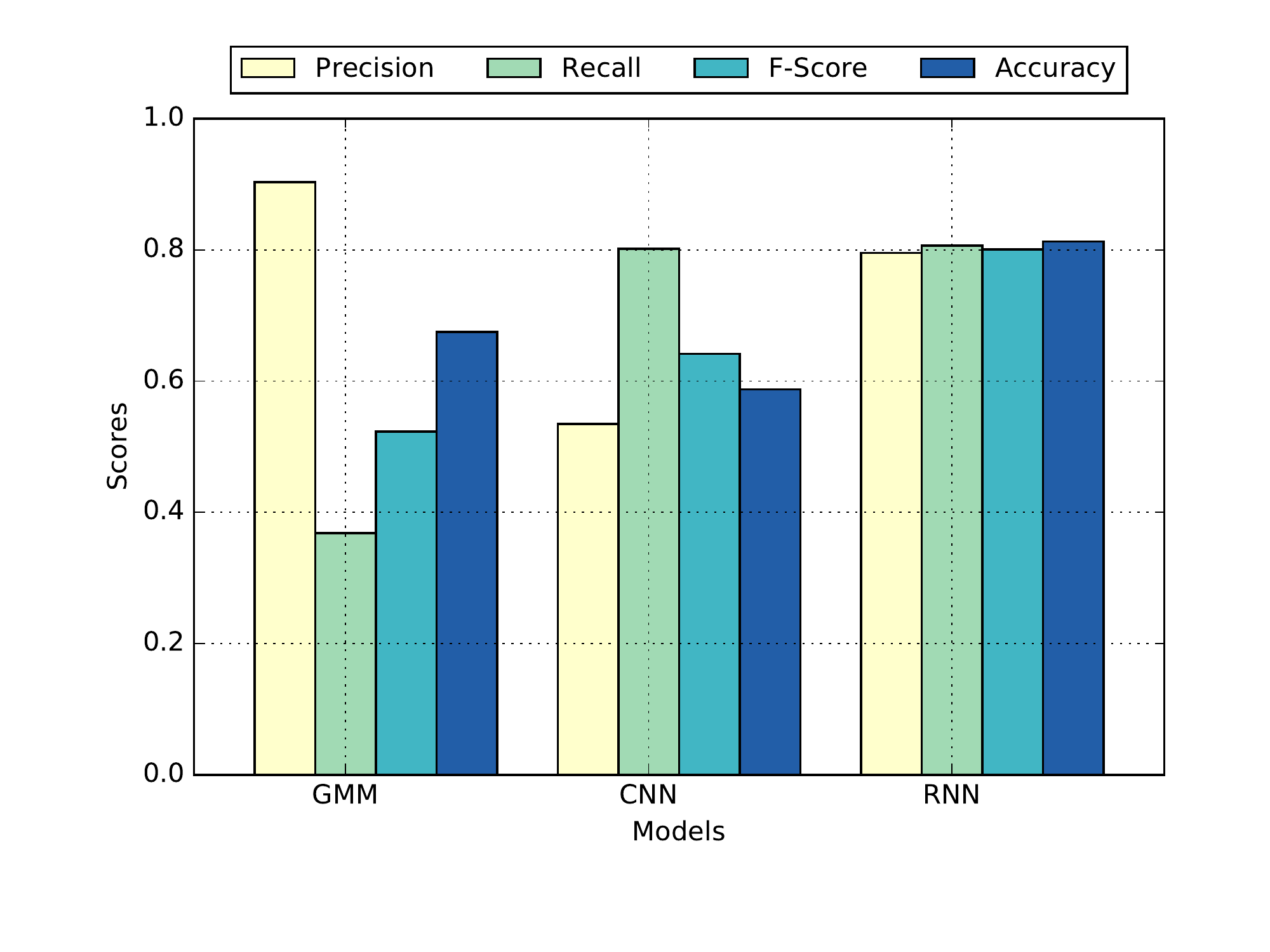}
		\caption{Detection performance for testing dataset. The RNN achieves the best performance on the training datasets with F-Score of 0.8009}
		\label{fig:eval_detection}
	\end{figure}

	We evaluated the detection performance of the drone using the proposed three models with the actual predicted period (Figure \ref{fig:label}), precision, recall, F-Score, and accuracy (Figure \ref{fig:eval_detection}). In our experiment, RNN achieves the best performance on the training datasets in terms of F-Score (RNN $>$ CNN $>$ GMM: 0.8009 $>$ 0.6415 $>$ 0.5232). Our RNN also shows the most balanced detection performance between precision and recall (0.7953, 0.8066). It is evident that our data augmentation is meaningful to remedy the shortage of the drone training dataset through this high-detection performance. Our CNN model is reported as the second best model in terms of F-Score. We note that it remains difficult to decide whether our CNN model outperforms GMM. Our CNN and GMM show a distinctly different tendency according in terms of precision (CNN, GMM: 0.5346 $<$ 0.9031) and recall (CNN, GMM: 0.8019 $>$ 0.3683). Our CNN shows the tendency to predict data as positive rather than negative. On the contrary, GMM treats most of the data as negative, thus it shows lower recall but higher precision. However, considering our detection label result (Figure \ref{fig:label}), GMM shows more accurate detection performance than statistics, but discontinuity in the positive prediction degrades the detection performance. This unstable consistency of positive prediction can be remedied by smoothing techniques. Therefore, in view of the operator, GMM can be regarded a more appropriate detection model to operate in practice. We also report the accuracy of these models, but do not consider it as important as the other measures. 
	
	Despite our diverse attempts we were unable to find CNN model architecture for previously proposed models. This could be attributed to the variation in audio data of the audio part unrelated to drone sound. In a real environment, we observe that the noticeable area affected by drone sound is small compared with the entire spectrogram image. Because of the fundamental mechanism of CNN, it is easily influenced by the other different areas of the spectrogram consisting of diverse environmental sound rather than focusing on drone sound only.
	
	\subsection{Testing: unseen types of data}
	The drawback of the machine-learning approach is the possibility of significant deterioration of detection performance when processing unlearned data. In this experiment, we aim to report degradation of detection performance for unseen types of data and improved understanding of the tendency of the proposed model. Our RNN still achieves the best performance in terms of F-Score (0.6984) with balanced precision and recall (0.5477, 0.9635). Interestingly, our report shows that the CNN model failed to classify the data, instead treating all data as positive. This misclassification could be caused by unseen highly noisy background sound that could not be distinguished from drone sound by the CNN model. According to this result, the CNN model is vulnerable to unseen noisy background data. Our GMM exhibits more accurate detection performance than CNN, but has a significantly decreased measure such that it would not be appropriate to operate in practice (0.3910 of F-Score).
	
	Our experiment with unseen data provides additional insight on the tendency of the proposed models for GMM to predict data as negative and the other models based on deep neural networks to predict data as positive. In our experiment, even introducing additional training data does not improve the GMM model significantly; however, RNN can improve their precision performance through the diverse background training dataset. Above all, this experiment confirmed that it is essential to collect diverse types of data for the target environment.
	
  \vspace{-0.3 in}
	
	\begin{figure}[ht!]
		\centering
		\includegraphics[scale=0.35]{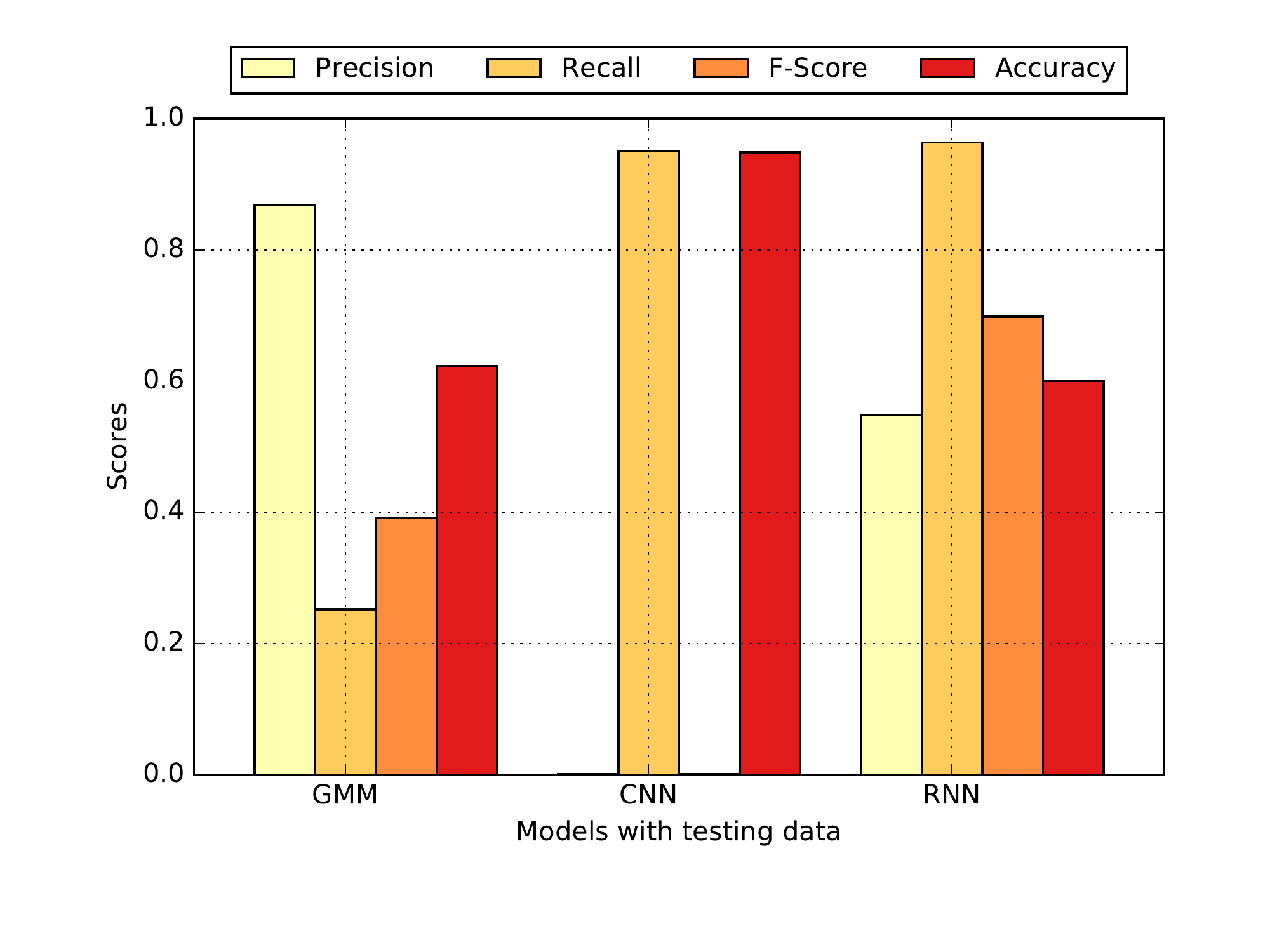}
		\caption{Performance comparison for unseen type data. The RNN still achieves the best performance with F-Score of 0.6984.}
		\label{fig:eval_seen}
	\end{figure}
	
	\subsection{Required cost versus Detection performance}
	In practice, the other factor that should be considered to operate the detection system is the cost required by the detection model. The two main factors determining the required cost for operating this system in practice are the processing time and the amount of input data for prediction. Many previous studies overlooked the practical constraint for improved detection performance, but it indeed may dominate the form of the system in a real environment, such as the difference in processing time between the GMM and DNN models \cite{sigtia2016automatic}. Statistically, we only measured the execution time for each process except for the additional time required to execute the Python program. According to the results, all three proposed models seem appropriate for application to real-time systems. The most time-consuming stage is feature engineering to create the MFCC vector, but it only takes 145 ms for a 1-minute audio clip. The execution time for data loading varies according to the target platform and the classification time does not adversely affect real-time system operation. For a fair comparison, we report the processing time for CNN and RNN without GPU usage. 
	
	However, we should note that the reported processing time is the minimal partial execution time only for each stage. In practice, the execution time of the detection algorithm varies according to the platform and is influenced to a greater extent by other costs related to program execution. Especially, if we plan to operate with a different programming language, then importing the Python program into a system programmed in another language would seriously deteriorate the execution time. Second, we note that models based on a deep neural network require a larger amount of data than GMM for optimal performance (240 ms $>$ 40 ms). This indicates that our actual initial detected time would increase as a function of the amount of input data. If we avoid importing the program or operating a low-performance embedded platform, the amount of input data can affect the initial detection time substantially.

	\begin{table}[ht!]
		\caption{Execution time for 1-minute audio clip}
		\label{table:eTime}
		\centering
		\begin{tabular}{|l | r|}
			\hline
			Process & Execution time (s) \\ \hline
			Data read from wav format &  0.0500 \\
			Feature engineering &  0.1451 \\
			Prediction: GMM & 0.0088 \\
			Prediction: CNN & 0.0473 \\ 
			Prediction: RNN & 0.0116 \\
			\hline
		\end{tabular}
	\end{table}
	

	\section{Conclusion}
	This paper presents our binary classification model that uses audio data to detect the existence of a drone. We configured the parameters for GMM and the network for CNN and RNN for our model. Then, we evaluated their detection performance in terms of the F-Score and the required cost for application to real-time systems in practice. In our experiment, the RNN model showed the best F-Score (0.8009) with 240 ms of audio input data. Our experiment also confirmed that the use of data augmentation to synthesize raw drone sound with diverse background sounds can alleviate the shortage of drone training data.
	
	The other main concern of our work was the influence on the detection performance of increasing drone distance. Because of practical constraints, we could not evaluate distances exceeding 150 m. In our experience, audio data recorded at distances further than 150 m do not display noticeable characteristics on the spectrogram with a na\''{i}ve recording with a single microphone. This was attributed to the drone sound exhibiting weakened characteristics in the spectrogram because it is covered by background data. The usage of multiple microphones with Beamforming, a signal processing technique used for filtering to achieve directional signal transmission, is expected to increase the maximum detection distance of our model. The other interesting future work would be utilization of the Generative Adversarial Network (GAN) to remedy the shortage of drone sound training data.
	
	
	\bibliographystyle{IEEEtran}
	
	\bibliography{./DD_EUSIPCO}

\begin{thebibliography}{10}
\providecommand{\url}[1]{#1}
\csname url@samestyle\endcsname
\providecommand{\newblock}{\relax}
\providecommand{\bibinfo}[2]{#2}
\providecommand{\BIBentrySTDinterwordspacing}{\spaceskip=0pt\relax}
\providecommand{\BIBentryALTinterwordstretchfactor}{4}
\providecommand{\BIBentryALTinterwordspacing}{\spaceskip=\fontdimen2\font plus
\BIBentryALTinterwordstretchfactor\fontdimen3\font minus
  \fontdimen4\font\relax}
\providecommand{\BIBforeignlanguage}[2]{{%
\expandafter\ifx\csname l@#1\endcsname\relax
\typeout{** WARNING: IEEEtran.bst: No hyphenation pattern has been}%
\typeout{** loaded for the language `#1'. Using the pattern for}%
\typeout{** the default language instead.}%
\else
\language=\csname l@#1\endcsname
\fi
#2}}
\providecommand{\BIBdecl}{\relax}
\BIBdecl

\bibitem{mezei2015drone}
J.~Mezei, V.~Fiaska, and A.~Moln{\'a}r, ``Drone sound detection,'' in
  \emph{Computational Intelligence and Informatics (CINTI), 2015 16th IEEE
  International Symposium on}.\hskip 1em plus 0.5em minus 0.4em\relax IEEE,
  2015, pp. 333--338.

\bibitem{busset2015detection}
J.~Busset, F.~Perrodin, P.~Wellig, B.~Ott, K.~Heutschi, T.~R{\"u}hl, and
  T.~Nussbaumer, ``Detection and tracking of drones using advanced acoustic
  cameras,'' in \emph{SPIE Security+ Defence}.\hskip 1em plus 0.5em minus
  0.4em\relax International Society for Optics and Photonics, 2015, pp.
  96\,470F--96\,470F.

\bibitem{mezei2016drone}
J.~Mezei and A.~Moln{\'a}r, ``Drone sound detection by correlation,'' in
  \emph{Applied Computational Intelligence and Informatics (SACI), 2016 IEEE
  11th International Symposium on}.\hskip 1em plus 0.5em minus 0.4em\relax
  IEEE, 2016, pp. 509--518.

\bibitem{mendis2016deep}
G.~J. Mendis, T.~Randeny, J.~Wei, and A.~Madanayake, ``Deep learning based
  doppler radar for micro uas detection and classification,'' in \emph{Military
  Communications Conference, MILCOM 2016-2016 IEEE}.\hskip 1em plus 0.5em minus
  0.4em\relax IEEE, 2016, pp. 924--929.

\bibitem{nguyen2016investigating}
P.~Nguyen, M.~Ravindranatha, A.~Nguyen, R.~Han, and T.~Vu, ``Investigating
  cost-effective rf-based detection of drones,'' in \emph{Proceedings of the
  2nd Workshop on Micro Aerial Vehicle Networks, Systems, and Applications for
  Civilian Use}.\hskip 1em plus 0.5em minus 0.4em\relax ACM, 2016, pp. 17--22.

\bibitem{barchiesi2015acoustic}
D.~Barchiesi, D.~Giannoulis, D.~Stowell, and M.~D. Plumbley, ``Acoustic scene
  classification: Classifying environments from the sounds they produce,''
  \emph{IEEE Signal Processing Magazine}, vol.~32, no.~3, pp. 16--34, 2015.

\bibitem{pohjalainen2011detection}
J.~Pohjalainen, T.~Raitio, and P.~Alku, ``Detection of shouted speech in the
  presence of ambient noise.'' in \emph{INTERSPEECH}, 2011, pp. 2621--2624.

\bibitem{mesaros2016tut}
A.~Mesaros, T.~Heittola, and T.~Virtanen, ``Tut database for acoustic scene
  classification and sound event detection,'' in \emph{Signal Processing
  Conference (EUSIPCO), 2016 24th European}.\hskip 1em plus 0.5em minus
  0.4em\relax IEEE, 2016, pp. 1128--1132.

\bibitem{zhang2015robust}
H.~Zhang, I.~McLoughlin, and Y.~Song, ``Robust sound event recognition using
  convolutional neural networks,'' in \emph{2015 IEEE International Conference
  on Acoustics, Speech and Signal Processing (ICASSP)}.\hskip 1em plus 0.5em
  minus 0.4em\relax IEEE, 2015, pp. 559--563.

\bibitem{cakir2016filterbank}
E.~Cakir, E.~C. Ozan, and T.~Virtanen, ``Filterbank learning for deep neural
  network based polyphonic sound event detection,'' in \emph{Neural Networks
  (IJCNN), 2016 International Joint Conference on}.\hskip 1em plus 0.5em minus
  0.4em\relax IEEE, 2016, pp. 3399--3406.

\bibitem{parascandolo2016recurrent}
G.~Parascandolo, H.~Huttunen, and T.~Virtanen, ``Recurrent neural networks for
  polyphonic sound event detection in real life recordings,'' in \emph{2016
  IEEE International Conference on Acoustics, Speech and Signal Processing
  (ICASSP)}.\hskip 1em plus 0.5em minus 0.4em\relax IEEE, 2016, pp. 6440--6444.

\bibitem{Grootel2009}
J.~K. M.W.W.~Grootel, T.C.~Andringa, ``{DARES-G1}: {D}atabase of {A}nnotated
  {R}eal-world {E}veryday {S}ounds,'' in \emph{Proceedings of the NAG/DAGA
  Meeting 2009}, 2009.

\bibitem{rakotomamonjy2015histogram}
A.~Rakotomamonjy and G.~Gasso, ``Histogram of gradients of time-frequency
  representations for audio scene classification,'' \emph{IEEE/ACM Transactions
  on Audio, Speech and Language Processing (TASLP)}, vol.~23, no.~1, pp.
  142--153, 2015.

\bibitem{sigtia2016automatic}
S.~Sigtia, A.~M. Stark, S.~Krstulovi{\'c}, and M.~D. Plumbley, ``Automatic
  environmental sound recognition: Performance versus computational cost,''
  \emph{IEEE/ACM Transactions on Audio, Speech, and Language Processing},
  vol.~24, no.~11, pp. 2096--2107, 2016.

\end{thebibliography}
	
\end{document}